\documentclass[a4paper, 10pt, conference, twocolumn]{ieeetran}      % Use this line for a4 paper

\IEEEoverridecommandlockouts                              % This command is only needed if 
\usepackage{graphics} % for pdf, bitmapped graphics files
\usepackage{epsfig} % for postscript graphics files
\usepackage{amsmath} % assumes amsmath package installed
\usepackage{amsthm}
\usepackage{amssymb}
\title{\LARGE \bf
Markovian Performance Model for Token Bucket Filter with Fixed and Varying Packet Sizes
}

\providecommand{\keywords}[1]
{
  \small	
  \textbf{\textit{Keywords---}} #1
}
				
\author{Henrik Schi{\o}ler \and John Leth \and Shibarchi Majumder  %<-this % stops a space
\thanks{Section for Control and Automation, Aalborg Universty, Aalborg, Denmark
        {\tt\small (henrik,jjl,sm)@es.aau.dk}}%
}

\begin{document}

\maketitle
\thispagestyle{empty}
\pagestyle{empty}
\newtheorem{thh}{Lemma}

%%%%%%%%%%%%%%%%%%%%%%%%%%%%%%%%%%%%%%%%%%%%%%%%%%%%%%%%%%%%%%%%%%%%%%%%%%%%%%%%
\begin{abstract}
We consider a token bucket mechanism serving a heterogeneous flow with a focus on backlog, delay and packet loss properties. Previous models have considered the case for fixed size packets, i.e. {\em one token per packet} with and M/D/1 view on queuing behavior. We partition the heterogeneous flow into several packet size classes with individual Poisson arrival intensities. The accompanying queuing model is a {\em full state} model, i.e. buffer content is not reduced to a single quantity but encompasses the detailed content in terms of packet size classes. This yields a high model cardinality for which upper bounds are provided. Analytical results include class specific backlog, delay and loss statistics and are accompanied by results from discrete event simulation. 
\end{abstract}

%%%%%%%%%%%%%%%%%%%%%%%%%%%%%%%%%%%%%%%%%%%%%%%%%%%%%%%%%%%%%%%%%%%%%%%%%%%%%%%%
\section{Introduction}
The token and leaky bucket mechanism have been applied widely as regulators for irregular traffic in packet switched computer networks, telecommunications networks and embedded systems \cite{Turner}, \cite{Swarna}. They may serve to provide tighter non-deterministic bounds in {\em Network Calculus} models \cite{Boudec} as well as well-defined bounds in communication topologies with {\em cyclic dependence} \cite{Schioler_2007}. Last but not least they may transform probabilistic uncertainties into non-deterministic uncertainties when applied to traffic flows under a probabilistic modeling regime, such as continuous time Markov models. In the latter case the token bucket mechanism in conjunction with the ingress flow characteristics provide the basis for queuing based performance models, whereas the token bucket parameters directly can be transformed to non-deterministic flow bounds of the egress flow to be applied downstream in e.g. network calculus performance modeling.\\
Our focus is the probabilistic performance modeling of the ingress buffer of a token bucket filter serving a compound Poisson arrival process with a discrete packet size distribution concentrated on a finite set. Previous token bucket models exist such as \cite{Jennings} where two {\em one token per packet} models are investigated; the {\em Cell Level} model and the {\em Token Level} model. Whereas the former constitutes a Markovian approximation the latter approaches the real system further by employing a discrete time Markov chain view point in terms of an M/D/1 queuing model. In \cite{Gebali} (ch. 8) Markovian approximate models are presented for a fixed packet length token bucket filter, for both single and burst arrivals. In this paper we give precise probabilistic models for a token bucket filter under both fixed and variable packet length assumptions. For the {\em fixed packet length} case we give a modified version of the Token Level model of \cite{Jennings}, which adopts the same observation instants right after token replenishment but assumes instant service in case of available tokens. We consider this model to be more precise than the model given in \cite{Jennings}, where the filter is modeled as a server. Whereas the M/D/1 model is known to reproduce time averages our model does not posses this quality. Therefore one needs to accompany discrete time results with time continuous output analysis to obtain precise results for waiting times and packet loss probabilities. In comparison with \cite{Gebali} we do not consider burst arrivals but variable packet lengths and our model does not suffer from the Markovian approximation found in \cite{Gebali}.\\
We first give algorithmic descriptions of the token bucket filter dynamics for the {\em fixed packet length} and {\em variable packet length} assumptions. The algorithmic descriptions are followed by probabilistic modeling sections for the {\em fixed packet length} and {\em variable packet length} cases. Hereafter a section provides the continuous time output analysis required to obtain precise results for waiting time and packet loss probabilities. Each theory-section is followed by a results section providing selected numerical results illustrating the virtues of the developed models in terms of available outputs and model flexibility.

\section{Token Bucket Algorithm}
\begin{figure}[h]
\begin{center}
\includegraphics[height=5cm]{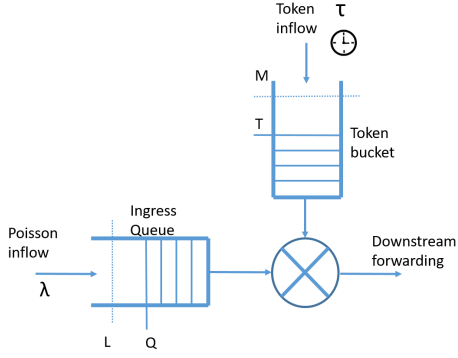}
\end{center}
\caption{Token bucket filter with fixed packet lengths}
\label{TB1}
\end{figure}

We assume periodic token replenishment, i.e. every $\tau$ time units a token is granted to the system, whereas at any time a packet may arrive to the ingress queue for potential entrance. We generally assume a {\em First Come First Served/Out} (FCFS/FIFO) discipline. The state of the system is at any time the number of tokens $T$ and the content of the ingress buffer. We denote the cumulatived backlog as $Q$. $T$ and $Q$ are limited to $M$ and $L$ respectively.  Packet length are given in units of tokens, i.e. how many tokens a packet consumes upon departure from the ingress buffer of the filter. For fixed packet lengths, as shown in figure \eqref{TB1} we generally assume packet lengths $1$, which leads to the following behavior
\begin{itemize} 
\item when a token arrives 
\begin{itemize}
\item if (the buffer is non-empty, i.e. $Q>0$) the buffer head is instantly\footnote{\label{fn1} {\em Instant} in this context means {\em in zero time}. {\em Clock-cycle} as well as {\em atomicity} details are abstracted away in this treatment.} removed\footnote{\label{fn2} Other than removal from the ingress buffer we do not consider the downstream fate of the departed package.} from the buffer and $Q=Q-1$.
\item else 
\begin{itemize}
\item if (the bucket is full, i.e. $T=M$) the token is discarded
\item else $T=T+1$
\end{itemize}
\end{itemize} 
\item when a packet arrives to the ingress buffer
\begin{itemize}
\item if (the buffer is full, i.e. $Q=L$) the packet is discarded
\item else
\begin{itemize}
\item if (there are available tokens, i.e. $T>0$) the packet is instantly removed from the system and $T=T-1$
\item else the packet is added to the buffer and $Q=Q+1$
\end{itemize}
\end{itemize}
\end{itemize} 

\begin{figure}[h]
\begin{center}
\includegraphics[height=5cm]{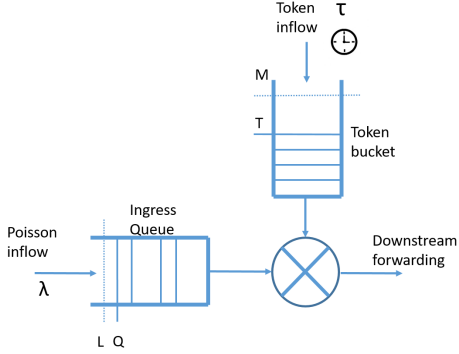}
\end{center}
\caption{Token bucket filter with variable packet lengths}
\label{TB2}
\end{figure}

If variable packet lengths are assumed (as shown in figure \eqref{TB2}), i.e. when packets consume a variable number of tokens upon transfer (removal) we have the following more complex dynamics
\begin{itemize} 
\item when a token arrives 
\begin{itemize}
\item if (the buffer is non-empty and $T+1$ is no less than the length of the buffer head $z_1$, i.e. $T+1 \geq z_l$) the buffer head is instantly removed from the buffer and $T=T+1-z_1$
\item else 
\begin{itemize}
\item if (the bucket is full, i.e. $T=M$) the token is discarded
\item else $T$ is incremented by 1
\end{itemize}
\end{itemize} 
\item when a packet of length $l$ arrives to the ingress buffer
\begin{itemize}
\item if (there is not enough buffer space available, i.e. $Q+l>L$) the packet is discarded
\item else
\begin{itemize}
\item if (the buffer is empty and there are enough available tokens, i.e. $T\geq l$) the packet is instantly removed from the system and $T=T-l$
\item else the packet is added to the buffer and $Q=Q+l$
\end{itemize}
\end{itemize}
\end{itemize} 

In the algorithm for variable packet sizes it is assumed that all packet sizes are less than the buffer length $L$. At a first glance it may seem, that some cases are left unresolved in the algorithms above. Like e.g. for fixed packet lengths, one might miss the case for arrival at a non-empty buffer with available tokens. In the following it is however proven, that if the buffer is initially empty such a state can never occur. Likewise, for variable packet lengths, the case for arrival to a buffer with insufficient space and available tokens seems unresolved. However a non-empty buffer means that a buffer head packet, already in the buffer, is stuck with insufficient tokens. Due to the FCFS service discipline tokens will not be available for the newly arrived packet. 

\section{Probabilistic Model for the Fixed Packet length case}
In the case for fixed packet lengths, the ingress buffer state is captured by a single non-negative integer $Q(t)$ obeying buffer length restrictions, i.e. $Q(t) \in \{0,..,L\}$, where $L$ is the buffer length. Likewise is the bucket state captured by a non-negative integer $T(t) \in \{0,..,M\}$, where $M$ is the bucket size. Two types of events; token replenishment and arrival to the ingress buffer change the state of the overall system. Token replenishments happen at instants $t=n\tau$ ($n \in \mathbb{N}$) and arrivals generally at instants $t \in \mathbb{R}_+$. We consider the four quantities $\{Q^-_n=Q(n\tau^-),Q^+_n=Q(n\tau^+),T^-_n=T(n\tau^-),T^+_n=T(n\tau^+)\}$ where the $^-$ superscript indicates state immediately before token replenishment at $t=n\tau$ and $^ +$ the state immediately after token replenishment when eventual token consumption has been accounted for. All in all, we may write the following dynamics at token replenishment
\begin{eqnarray}
Q^+_n &=& \max\{0,Q^-_n-1\} \nonumber \\
T^+_n &=& T^-_n  \ \ \mbox{for} \ \ Q^-_n>0 \nonumber \\
T^+_n &=& \min\{M,T^-_n+1\}  \ \ \mbox{for} \ \ Q^-_n=0 \label{token}
\end{eqnarray}
and upon arrival for $t \in [(n-1)\tau, n\tau)$.
\begin{eqnarray}
Q(t^+) &=& Q(t^-) \ \ \mbox{for} \ \ T(t^-)>0 \nonumber \\
Q(t^+) &=& \min\{L,Q(t^-)+1\} \ \ \mbox{for} \ \ T(t^-)=0 \nonumber \\
T(t^+) &=& \max\{0,T(t^-)-1\}  \label{arrival} 
\end{eqnarray}
where $^-$ indicates state immediately before arrival and  $^+$ immediately after. This yields for the entire period $[(n-1)\tau, n\tau)$
\begin{eqnarray}
Q^-_{n} &=& \max\{0,\min\{L,Q^+_{n-1}+a_n-T^+_{n-1}\}\} \nonumber \\
T^-_{n} &=& \max\{0,T^+_{n-1}-Q^+_{n-1}-a_n\}  \label{entire}
\end{eqnarray}
where $a_n$ is an independent random variable accounting for Poisson arrivals within $[(n-1)\tau, n\tau)$. The following lemma, which is commonly known and proven in appendix, allows a convenient change of coordinate

\begin{thh}
$Q(0)T(0)=0$ implies $Q(t)T(t)=0$ identically for all $t>0$.
\end{thh}

\subsection{Change of coordinate}
Now assuming $Q(t)=0$, $Q(t)T(t)=0$ can be assumed identically or in turn either $Q(t)=0$ or $T(t)=0$ indefinitely. Therefor the entire state of the token bucket filter is appropriately captured in the state variable $K(t)=Q(t)-T(t) \in \{-M,..,L\}$, such that $Q(t)=\max\{0,K(t)\}$ and $T(t)=-\min\{0,K(t)\}$.\\
Dynamics for $K(t)$ can be expressed at token replenishment, as (shown in appendix)
\begin{equation}
K^+_n = \max\{-M,K^-_n-1\} \label{Ktoken}
\end{equation}
At arrival a similar analysis applies (as shown in appendix) to yield
\begin{equation}
K^-_{n} = \min\{L,K^+_{n-1}+a_n\} \label{Kentire} 
\end{equation}
Combining \eqref{Ktoken} and \eqref{Kentire} gives
\begin{equation}
K^+_n = \max\{-M,\min\{L,K^+_{n-1}+a_n\}-1\} \label{Kcomb}
\end{equation}
Defining $S(t)=K(t)+M$ yields
\begin{equation}
S^+_n = \max\{0,\min\{L+M,S^+_{n-1}+a_n\}-1\} \label{Scomb}
\end{equation}
Since $\{a_n\}$ is an i.i.d. sequence $\{S^+_n\}$ is a discrete time Markov chain (DTMC). The {\em token level} model of \cite{Jennings} likewise yields a DTMC following M/D/1/L+M dynamics. M/D/1/L+M dynamics however follow
\begin{eqnarray}
S^+_n &=& \max\{0,\min\{L+M,S^+_{n-1}+a_n\}-1\} \nonumber \\
& &\mbox{for} \ \ S^+_{n-1}>0 \nonumber \\
&=& \max\{0,\min\{L+M,a_n\}\} \ \ \mbox{for} \ \ S^+_{n-1}=0 \label{SDTMC}
\end{eqnarray}
which is clearly not identical to \eqref{Scomb}. We refer to the dynamics given by \eqref{Scomb} as {\em Periodic Transfer}, since it reflects the difference to M/D/1 service, that an packet arriving to an empty queue does not immediately receive service but has to wait for a token to arrive. Upon token arrival, processing the packet is comprised by instantaneous transfer out of the input buffer. Defining 
\begin{equation}
P_n=[P(S^+_n=0) \ P(S^+_n=1) \ .. \ P(S^+_n=L+M-1)] \nonumber
\end{equation}
gives according to \eqref{Scomb}
\begin{equation}
P_n=P_{n-1} {\bf H} \nonumber
\end{equation}
which can be solved under stationarity, i.e. $\pi^S = \pi^S {\bf H}$ and subsequently transformed according to
\begin{eqnarray}
P(Q^+_n=0)&=&P(S^+_n \leq M) \nonumber \\
P(Q^+_n=j>0)&=&P(S^+_n = j-M) \nonumber
\end{eqnarray} 
such that, $\pi^Q(0) = \sum_{k=0}^M \pi^S(k)$ and $\pi^Q(k>0) = \pi^S(k-M)$

\section{Probabilistic Model for variable packet lengths}
In the case for variable packet lengths we model the arrival process as a compound Poisson process $C$, where the size of the arrived packet is measured in integer multiples of tokens, i.e. each arrived packet will consume a number of tokens corresponding to its size, when transferred out of the ingress buffer. Since especially dynamics at token replenishment depends directly on the size of the packet heading the input buffer, the model should somehow reflect this aspect of system state. Mean field approximations could be applied drawing packet size randomly independent of history at every token replenishment instant according to stationary distributions. This however would create a {\em causality dilemma} since stationary distributions, at the same time, would be both inputs to and outputs from the model. Although this dilemma might be solved e.g. iteratively, it would be hard to obtain error bounds for such a model and it would not provide more involved higher order statistics as would be available from a {\em full state model}. Thus we proceed with a full state model, where state would comprise input buffer content in terms of buffered packet sizes as well as the number of stored tokens.\\

System state $x$ formally evolves within ${\cal X}=\{0,..,M\} \times Z^*_L = \{X_1,..,X_N\}$, $Z^*_L \subseteq Z^*$  where $Z$ is the finite set of packet lengths $\{l_1,..,l_W\}$, $Z^*$ is the language of strings ${\bf z}$ over $Z$ including the empty string $\epsilon$ and $Z^*_L$ is the subset such that string lengths obey $|{\bf z}|\leq L$, i.e.
\begin{equation}
Z^*_L=\{{\bf z} \in Z^* \ | \ |{\bf z}| \leq L\} \nonumber
\end{equation}
The interpretation of formal system state is fairly immediate, e.g. $(T,{\bf z})=(T,z_1,z_2,..,z_l)$ denotes the state, where $T$ tokens are stored and the buffered packets have lengths $z_1,z_2,..,z_l$ where $z_1$ is the length of the buffer head and $z_l$ indicates tail.\\
As above, system state is observed immediately after token replenishment where dynamics are as follows for a non-empty buffer:
{\small
\begin{eqnarray}
T_n^+,z_{1,n}^+,z_{2,n}^+,..,z_{l-1,n}^+ &=& T_n^--z_{1,n}^-+1,z_{2,n}^-,z_{3,n}^-,..,z_{l,n}^- \nonumber \\
& & \ \mbox{for} \ \ T_n^--z_{1,n}^-+1\geq 0 \nonumber \\ 
T_n^+,z_{1,n}^+,z_{2,n}^+,..,z_{l,n}^+ &=& \min\{M,T_n^-+1\},z_{1,n}^-,z_{2,n}^-,..,z_{l,n}^- \nonumber \\
& & \ \mbox{for} \ \ T_n^--z_{1,n}^-+1< 0 \label{dyn1}
\end{eqnarray}
}
where the former equation captures token bucket dynamics, when (after token replenishment) there are enough tokens available to transfer the heading packet out of the ingress buffer and the latter the opposite case. For an empty buffer we have
\begin{equation}
T_n^+,{\bf z}_n^+=\min\{M,T_n^-+1\},\varepsilon \label{dyn2}
\end{equation}
For packet arrival at instant $t$ the following dynamics define token bucket mechanics for the nonempty case
\begin{equation}
T(t^+),{\bf z}(t^+)=T(t^-),{\bf z}(t^-) c(t) \ \ \mbox{for} \ \ |{\bf z} \ c(t)| \leq L \label{dyn3}
\end{equation}
where $c(t)$ is the jump of the input compound Poisson process at time $t$, (i.e. $c(t)=C(t)-\lim_{\tau \rightarrow t^-} C(\tau)$ and $|{\bf z}|$ denotes the total backlog (in units of tokens) for buffer content ${\bf z}$.  Empty buffer dynamics are 

\begin{eqnarray}
T(t^+),{\bf z}(t^+) &=& T(t^-)-c(t),\varepsilon \nonumber \\
& & \ \mbox{for} \ \ T(t^-)-c(t)+1 \geq 0 \nonumber \\ 
T(t^+),{\bf z}(t^+) &=& T(t^-),c(t) \nonumber \\
& & \ \mbox{for} \ \ T(t^-)-c(t)+1 < 0 \label{dyn4}
\end{eqnarray}

\subsection{Probabilistic Modeling}
We proceed by transforming the functional dynamics for the variable size case into a probabilistic model. First we enumerate system states, i.e. $x \in {\cal X}=\{0,..,M\} \times Z^* = \{X_1,..,X_N\}$, where the cardinality $N=(M+1) |Z^*|$ and $|Z^*|$ is the cardinality of the set of $Z^*$.\\ 
We generally denote dynamics at token replenishment as
\begin{equation}
x_n^+ = \Gamma(x_n^-) \nonumber
\end{equation}
such that probabilistically
\begin{equation}
P(x_n^+ = X_j) = \sum_{k=1}^N I_{\Gamma(x_k^-)=X_j} P(x_n^- = X_k) \nonumber
\end{equation}
Between token replenishment system state evolves according to a continuous time Markov chain (CTMC) characterized by constant transition rates $\lambda_{i,j}$ for transition between state $x_i$ and $x_j$. Let the overall arrival rate of the compound Poisson process $C$ be $\lambda$ and the distribution over $Z$ be $\{u_1,..,u_W\}$. Then for ${\bf z} \neq \epsilon$ 

\begin{eqnarray}
X_j&=&T,{\bf z} \ l_k \nonumber \\
X_i&=&T,{\bf z} \nonumber
\end{eqnarray}
$\lambda_{i,j}=u_k \lambda I_{|{\bf z} l_k| \leq L}$, where $|{\bf z}|=\sum_i z_i$ denotes buffer occupancy for content encoded by ${\bf z}$ (not to be confused with string length).\\
For an empty buffer
\begin{eqnarray}
X_j&=&T-l_k,\epsilon  \nonumber \\ 
X_i&=&T,\epsilon \nonumber 
\end{eqnarray} 
$\lambda_{i,j}=u_k \lambda I_{T-l_k \geq 0}$ as well as

\begin{eqnarray}
X_j&=&T,l_k \nonumber \\
X_i&=&T,\epsilon \nonumber 
\end{eqnarray} 
$\lambda_{i,j}=u_k \lambda I_{T-l_k < 0}$.\\

Defining $P(t)=[P(x(t)=X_1),..,P(x(t)=X_N)]$ allows for a compact description of the probabilistic dynamics above. At token replenishment we have
\begin{equation}
P_n^+ = P_n^- {\bf H} \nonumber
\end{equation}
and between
\begin{equation}
\frac{d}{dt} P(t) = P(t) {\bf Q} \label{diffeq1}
\end{equation}
where ${\bf H}$ is an $N \times N$ transition matrix and ${\bf Q}$ is an $N \times N$ rate matrix with $Q_{i,j}=\lambda_{i,j}$. Solving \eqref{diffeq1} between token replenishment instants $n \tau$ and $(n+1) \tau$ yields
\begin{equation}
P_{n+1}^-=P_n^+ exp({\bf Q} \tau) \label{expm1}
\end{equation}
and further
\begin{equation}
P_{n+1}^+ = P_{n+1}^- {\bf H} = P_n^+ exp({\bf Q} \tau) {\bf H}  \label{iteration_full}
\end{equation}

Now $G=exp({\bf Q} \tau) {\bf H}$ is a {\em stochastic} matrix, so a unique limiting solution $\pi$ to $\{P_n^+\}$ exists for $G$ being {\em irreducible}, i.e. there is an $n \leq N$ such that $G^n$ has only non-zeros entries. Irreducibility is equivalent to the existence of a sequence, with positive probability, of at most $N$ token replenishment periods, taking system state from any element in the state space to any other element. We leave it to the reader to assure himself, that this is the case for the presented model. Under the same assumptions $\pi$ comprises a consistent estimator for discrete time averages, i.e. $\lim_{n \rightarrow \infty} \frac{1}{n} \sum_{i=1}^n I_{x_n^+ \in A}$, where $A \subseteq {\cal X}$.

\subsection{Partition of State Space}
The cardinality of the state space ${\cal X}$ is potentially very large. Computing $exp({\bf Q} \tau)$ may therefore be of tremendous complexity. In order to reduce computational complexity a partition of ${\cal X}$ is suggested such that continuous time dynamics evolve almost independently among subdivisions. Define $P^{\varepsilon}_t$ by
\begin{equation}
P^{\varepsilon}_t=[P(x(t)=0,\varepsilon),P(x(t)=1,\varepsilon),..,P(x(t)=M,\varepsilon)] \nonumber
\end{equation}
i.e. the distribution over all empty buffer states. Next, for non-empty buffer
\begin{equation}
P^T_t=[P(x(t)=T,{\bf z}_1),P(x(t)=T,{\bf z}_2),..,] \nonumber
\end{equation}
where the ordering $Z^*=\{{\bf z}_0,{\bf z}_1,..\}$ is lexicographic, i.e. ${\bf z}_0 = \varepsilon$.\\
A non-empty buffer between token replenishments is always a result of a buffer head packet too large for the available tokens, i.e. ${\bf z}=z_H {\bf z}_T$ and $T<z_H$. Therefore, there can be no {\em direct probabilistic flow} (between token replenishments) from states $T,{\bf z}$ to states $T',{\bf z}'$ for $T' \neq T > 0$ and ${\bf z} \neq \varepsilon$. This includes the special case for ${\bf z}' = \varepsilon$. Therefore a sub-stochastic matrix ${\bf Q}^{\varepsilon}$ exists such that\\
\begin{equation}
\frac{d}{dt} P^{\varepsilon}_t = P^{\varepsilon}_t {\bf Q}^{\varepsilon} \label{diffeq2}
\end{equation}
where 
\begin{eqnarray}
{\bf Q}^{\varepsilon}_{T,T-l_k}&=&u_k \lambda I_{T-l_k \geq 0} \nonumber \\
{\bf Q}^{\varepsilon}_{T,T}&=&-\lambda \sum_k u_k = -\lambda \nonumber 
\end{eqnarray}

as well a sub-stochastic matrix ${\bf Q}'$ and a matrix ${\bf B}^T$ such that
\begin{equation}
\frac{d}{dt} P^T_t = P^T_t {\bf Q}' +  P^{\varepsilon}_t {\bf B}^T \label{diffeq3}
\end{equation}
where 
\begin{eqnarray}
{\bf Q}'_{i,j}&=&u_k \lambda I_{|{\bf z}_i l_k| \leq L} \ \ \mbox{for} \ \ {\bf z}_j={\bf z}_i l_k \nonumber \\
{\bf Q}'_{i,i}&=&-\sum_k u_k \lambda I_{|{\bf z}_i l_k| \leq L} \nonumber
\end{eqnarray}
and
\begin{eqnarray}
{\bf B}^T_{T,p}&=&u_k \lambda I_{T-l_k < 0} \ \ \mbox{for} \ \ {\bf z}_p = l_k  \nonumber
\end{eqnarray}
  
Concatenating probability vectors, i.e. $P^{T'}_t=[P^{\varepsilon}_t \ P^T_t]$ leads to
\begin{equation}
\frac{d}{dt} P^{T'}_t = P^{T'}_t {\bf \Gamma}^T \label{diffeq4}
\end{equation}
where

\[
{\bf \Gamma}^T=\begin{bmatrix}
    {\bf Q}^{\varepsilon}       & {\bf B}^T \\
     {\bf 0}  & {\bf Q}'
\end{bmatrix}
\]
with a closed form solution
\begin{equation}
P^{T'}_{t+\tau} = P^{T'}_t exp({\bf \Gamma}^T \tau) \label{diffsolve}
\end{equation}
Now the solution in \eqref{diffsolve} has to be computed for all $T \in \{0,..,M\}$, where $exp({\bf \Gamma}^T \tau)$ is an $(M+N) \times (M+N)$ matrix, i.e. altogether $(M+1)(M+N)^2$ entries. On the other hand, solving \eqref{expm1} involves $exp({\bf Q} \tau)$ which is an $N \times N = (M+1)N\times (M+1)N$ matrix with $(M+1)^2 N^2$ entries. Since in almost all cases $1<<M<<N$ we have approximately in the former case $M N^2$ and in the latter $M^2 N^2$ and the computational benefit of state space partitioning clearly shows.

\section{Bounding state space cardinality}
The state space ${\cal X}=\{0,..,M\} \times Z^*_L=\{X_1,..,X_N\}$, where $Z^*_L \subseteq Z^*$, where
\begin{equation}
Z^*_L=\{{\bf z} \in Z^* \ | \ |{\bf z}| \leq L\} \nonumber
\end{equation}
We define the function $B:Z \times {\mathbf N} \rightarrow {\mathbf N}$ by
\begin{equation}
B(l,L)=\# \{{\bf z} \in Z^* \ | \ |l {\bf z}| \leq L\} \nonumber
\end{equation} or less formally the number of buffer states with a buffer head of length $l$. 
We readily identify
\begin{equation}
\# Z^*_L=\sum_{l \in Z} B(l,L)  \label{R} \nonumber
\end{equation}
It is proven recursively (in appendix), that
\begin{equation}
\# Z^*_L = (\#Z^{\frac{1}{\min\{ Z\}}})^L \label{Bbar}
\end{equation}
revealing a high cardinality for a small value of $\min\{ Z\}$ and vice versa. Intuitively large minimum packets do not allow high packet counts and therefore limit the cardinality.  

\subsection{Example}
We consider the cases $L=\{3,..,10\}$ and $Z=\{1,2,3,4\}$ or $Z=\{3,4,5,6\}$

\begin{figure}
\begin{center}
\includegraphics[height=5cm]{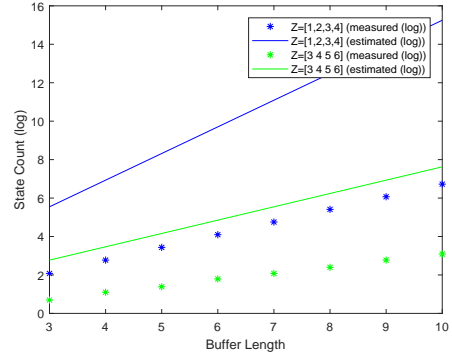}
\end{center} 
\caption{Estimated and measured state count.}
\label{statecount}
\end{figure}
Estimated state counts are computed based on \eqref{R} and \eqref{Bbar} and as comparison found through enumeration. Enumeration is performed over through lexicographic ordering and simultaneous counting of $Z^*_L$. Results are given logarithmically in figure \eqref{statecount}. Even though stimates capture the dependence of the growth on $\min\{ Z\}$, $\#Z^*_L$ is itself highly over-estimated. On the other hand, it is good news for the applicability of the method itself, that even for the situation with $Z=\{1,2,3,4\}$, where $\#Z^*_7$ is estimated to 4E6, $Z^*_7$ comprises only 833 states when counting.

\section{Numerical Validation}
\begin{figure}[htb]
\begin{center}
\includegraphics[height=5cm]{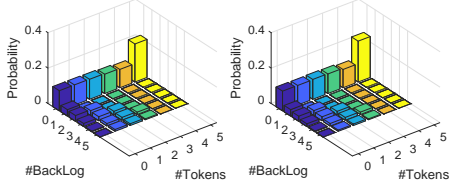}
\end{center} 
\caption{Simulation and numeric results for $T=1$, $\lambda=0.5$, $M=5$ and $L=5$ (Simulation to the left).}
\label{sim1}
\end{figure}

\begin{figure}[htb]
\begin{center}
\includegraphics[height=5cm]{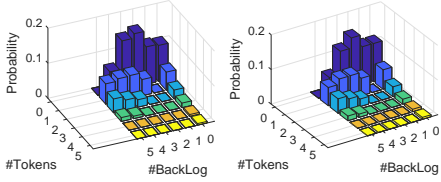}
\end{center} 
\caption{Simulation and numeric results for $T=1$, $\lambda=1$, $M=5$ and $L=5$ (Simulation to the left).}
\label{sim2}
\end{figure}

\begin{figure}[htb]
\begin{center}
\includegraphics[height=5cm]{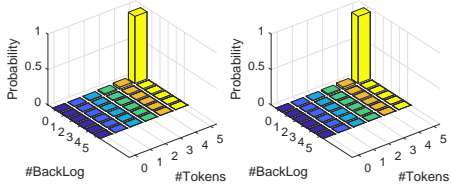}
\end{center} 
\caption{Simulation and numeric results for $T=1$, $\lambda=0.25$, $M=5$ and $L=5$ (Simulation to the left).}
\label{sim3}
\end{figure}

For validation numerical results are provided through two comparative approaches; numerical iteration over equation \eqref{iteration_full} is well as through stochastic simulation of equations \eqref{dyn1} through \eqref{dyn4}  with the TrueTime \cite{Truetime} toolbox for MATLAB SimuLink. To provide overview we present aggregate statistic, i.e. the joint distribution of pairs $(T,|{\bf z}|)$ - that is, available tokens and buffer occupancy. Figure \eqref{sim1} shows results for $T=1$, $\lambda=0.5$, $M=5$, $L=5$, $Z=\{1,2,3,4\}$ and $u=\{u_i\}=\{4,3,2,1\}/10$. Notice that $\lambda=\frac{T}{<Z,u>}$ illustrating the situation for $100\%$ load. 
Figure \eqref{sim2} shows results for the same settings except $\lambda=1$. That is, a highly loaded situation.
Figure \eqref{sim3} shows results for the same settings except $\lambda=0.25$. That is, a lightly loaded situation.
Inspecting results reveals an almost identical match between numerical and simulation results. On this basis we consider the model and the solver implementation validated.

\section{Continuous time results}
For many relevant performance statistics, results for observation instants $T^+_n$ may not be adequate, since the statistics for observation instants may not carry over to {\em random arrivers}. Thus, we need to convert DTMC results to time averages. Consider any subset $A \subseteq {\cal X}$ of the statespace. Then, the time averages $E_A$ of interest would be
\begin{equation}
E_A=\lim_{t \rightarrow \infty} \frac{1}{t} \int_0^t I_{x(\eta) \in A} d\eta \nonumber
\end{equation}
under stationarity and assuming ergodicity we may find $E_A$ as
\begin{eqnarray}
E_A&=&E[\frac{1}{\tau} \int_0^{\tau} I_{x(\eta) \in A} d\eta] \nonumber \\
&=&\frac{1}{\tau} \int_0^{\tau} E[I_{x(\eta) \in A}] d\eta \nonumber \\ 
&=&\frac{1}{\tau} \int_0^{\tau} P(x(\eta) \in A) d\eta \label{time_cont1}
\end{eqnarray}
where expectation- and probability-operators $E$ and $P$ are understood to be under stationarity conditions, i.e. $\{P_n^+\}=\pi$.\\
The limiting solution $\pi$ to \eqref{iteration_full} may be partitioned as
\begin{equation} 
\pi=[\pi_0 \ \pi_1 \ .. \ \pi_M]
\end{equation}
where
\begin{equation} 
\pi_T=[\pi_T^1 \ \pi_T^2 \ .. \ \pi_T^{|Z^*|}]
\end{equation}

Thus, the solution $P^{T'}_t$ to \eqref{diffeq4} limits to $[[\pi^1_0 ..\pi^1_M]  \pi_T]$ where $[\pi^1_0 ..\pi^1_M]$ comprise all stationary empty-buffer state probabilities. Therefore we conveniently partition $E_A$ as (where $k$ can be any index in $\{0,..M\}$
%For simplicity we assume $A \cap \{x = T,{\bf z} \ | \ {\bf z} = \varepsilon\} = \emptyset$ and $T,{\bf z} \in A \Leftrightarrow  T',{\bf z} \in A$. 
\begin{eqnarray}
E_A&=&\sum_{l=0}^M \frac{1}{\tau} \int_0^{\tau} P(x(\eta) \in A \ \wedge \ T(t)=l \ \wedge \ {\bf z} \neq \ \varepsilon) \nonumber \\
&+&P(x(\eta) \in A \ \wedge \ T(t)=l \ \wedge \ {\bf z} = \varepsilon) d\eta \nonumber \\
&=&\sum_{l=0}^M \frac{1}{\tau} \int_0^{\tau} P^{l'}_{\eta} {\mathbf 1}_{l,A^{\bar{\epsilon}}}  d\eta
+\frac{1}{\tau} \int_0^{\tau} P^{k'}_{\eta} {\mathbf 1}_{0,A^{\epsilon}}  d\eta \nonumber \\
&=&\sum_{l=0}^M \frac{1}{\tau} \int_0^{\tau} [[\pi^1_0 ..\pi^1_M]  \pi_l] exp(\Gamma^l \eta) {\mathbf 1}_{l,A^{\bar{\epsilon}}}  d\eta \\
&+&\frac{1}{\tau} \int_0^{\tau} [[\pi^1_0 ..\pi^1_M]  \pi_k] exp(\Gamma^k \eta) {\mathbf 1}_{0,A^{\epsilon}}  d\eta \label{expmint}
\end{eqnarray}
where  $A^{\epsilon}=A \cap \{x = T,{\bf z} \ | \ {\bf z} = \varepsilon\}$, $A^{\bar{\epsilon}}=A \setminus A^{\epsilon}$ and ${\mathbf 1}_{l,B}$ is an indicator column vector, i.e. ${\mathbf 1}_{l,B}^j = I_{(l,{\bf z}_j) \in B}$ (recall $Z^*=\{{\bf z}_1,..,{\bf z}_{|Z^*|}\}$).
%The limiting solution $\pi$ to \eqref{iteration_full} may be partitioned as
%\begin{equation} 
%\pi=[\pi_0 \ \pi_1 \ .. \ \pi_M]
%\end{equation}
%where
%\begin{equation} 
%\pi_T=[\pi_T^1 \ \pi_T^2 \ .. \ \pi_T^{|Z^*|}]
%\end{equation}
%where $\pi_T^j$ is the limiting probability for $x_n^+ = T,{\bf z}_j$. Thus we may manifest \eqref{EA} as
%\begin{eqnarray}
%E_A&=&\sum_{l=0}^M \frac{1}{\tau} \int_0^{\tau} P^{l'}_{\eta} {\mathbf 1}_A  d\eta \nonumber \\
%&=&\sum_{l=0}^M \frac{1}{\tau} \int_0^{\tau} [[\pi^1_0 ..\pi^1_M]  \pi_l] exp(\Gamma^l \eta) {\mathbf 1}_A  d\eta \label{expmint}
%\end{eqnarray}
%where it should be noticed that $[\pi^1_0 ..\pi^1_M]$ comprise all stationary empty-buffer state probabilities. 

Since $\Gamma^l$ is generally not non-singular the integral in \eqref{expmint} may require decomposition techniques like Jordan decomposition \cite{Jordan}. We leave the general computational complexity issues outside the scope of this work and compute the integral numerically.\\

\subsection{Example - loss probabilities}
\begin{figure}[htb]
\begin{center}
\includegraphics[height=5cm]{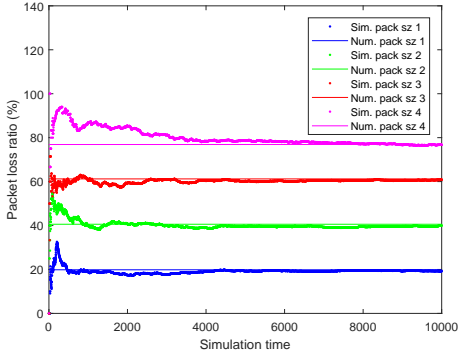}
\end{center} 
\caption{Simulation and numeric results for individual packet loss ratios - high load.}
\label{sim4}
\end{figure}

We choose packet drop/loss ratios ${\cal L}(l)$ for specific packet sizes $l$ as the targeted statistics. The indicator vector above is therefore given as $I_A^j=\{|{\bf z}_j|>L-l\}$. Simulation and numerical results for packet sizes $Z=\{1,2,3,4\}$ are shown for the heavily loaded case, i.e. $\lambda=5$ in figure \eqref{sim4}

\begin{figure}[htb]
\begin{center}
\includegraphics[height=5cm]{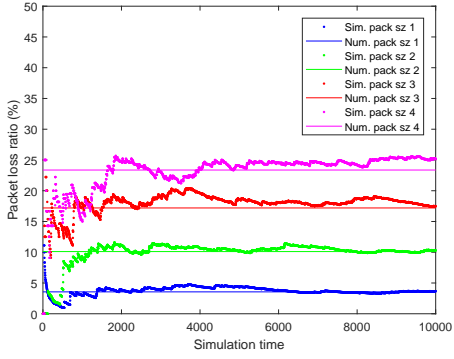}
\end{center} 
\caption{Simulation and numeric results for individual packet loss ratios - light load.}
\label{sim5}
\end{figure}

whereas for the lightly loaded case ($\lambda=0.25$) results are shown in figure \eqref{sim5}
In both the heavy- and light-load cases it numerical and simulation results show precise compliance.

\subsection{Example - waiting time}

For average waiting time $E_W(k)$ for packets of size $l_k$ we apply the {\em Little}-identity \cite{Kleinrock} to results for the backlog time average $E_Q(k)$. We define for a backlog configuration ${\bf z}$ the function $Qc(k,{\bf z})$ yielding the number of appearances of packets with length $l_k$ in ${\bf z}$. Then the backlog time average $E_Q(k)$ is found by
\begin{eqnarray}
E_Q(k)&=&E[\frac{1}{\tau} \int_0^{\tau} Qc(k,{\bf z}(\eta)) d\eta] \nonumber \\
&=&\frac{1}{\tau} \sum_{l=0}^M \sum_{j=1}^{|Z^*|} \int_0^{\tau} P(x(\eta)=l,{\bf z_j}) Qc(k,{\bf z}_j) d\eta \nonumber \\
&=&\frac{1}{\tau} \sum_{l=0}^M \sum_{j=2}^{|Z^*|} \int_0^{\tau} P(x(\eta)=l,{\bf z_j}) Qc(k,{\bf z}_j) d\eta \nonumber \\
&=&\frac{1}{\tau} \sum_{l=0}^M \sum_{j=2}^{|Z^*|} [[\pi^1_0 ..\pi^1_M]  \pi_l]  \nonumber \\
& &\ \ \ \ \ \ \int_0^{\tau} exp(\Gamma^l \eta) d\eta \ Qc(k,{\bf z}_j) \ {\mathbf 1}_{j+M} \nonumber 
\end{eqnarray}
With application of Little's identity we obtain
\begin{equation}
E_W(k)=\frac{E_Q(k)}{(1-{\cal L}(l)) \lambda u_k} \nonumber
\end{equation}

\begin{figure}[htb]
\begin{center}
\includegraphics[height=5cm]{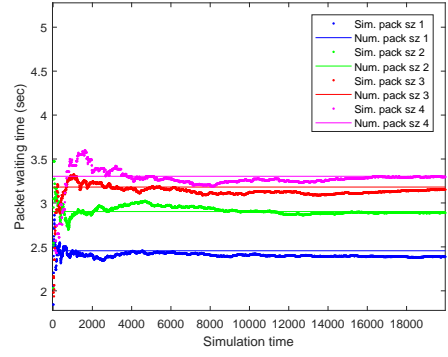}
\end{center} 
\caption{Simulation and numeric results for individual packet waiting time - high load.}
\label{sim6}
\end{figure}

Simulation and numerical results for packet sizes $Z=\{1,2,3,4\}$ are shown for the heavily loaded case, i.e. $\lambda=5$ in figure \eqref{sim6}

\begin{figure}[htb]
\begin{center}
\includegraphics[height=5cm]{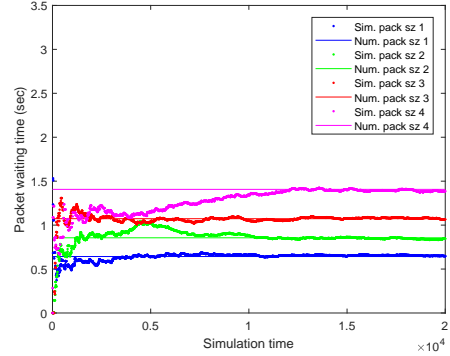}
\end{center} 
\caption{Simulation and numeric results for individual packet waiting time - light load.}
\label{sim7}
\end{figure}

whereas for the lightly loaded case ($\lambda=0.25$) results are shown in figure \eqref{sim7}
In both the heavy- and light-load cases it numerical and simulation results show compliance within acceptable range.

\section{Conclusion}
A probabilistic model for token bucket mechanisms serving both homogeneous and heterogeneous flow with a focus on backlog, delay and packet loss properties has been developed. The developed model is a {\em full state} model encompassing detailed buffer content in terms of packet size classes. This yields a high model cardinality for which upper bounds are provided. We provide analytical results for class specific backlog, delay and loss statistics. Illustrative examples are provided on which theoretical results are compared with results from discrete event simulation. Results show precise conformance in all cases. application of the developed model for computing central performance statistics holds several advantages over the use of discrete event simulation; namely discrete event simulation, for all the presented cases, requires significant higher run time and it is difficult to predict the needed run time for a required level of precision. We predict that thorough {\em output analysis} of discrete event result would require a model almost equally complex as the developed one.\\
Therefore we conclude, that the developed model framework constitutes a solid basis for performance evaluation of token bucket filters serving heterogeneous flows.\\
Directions for future research include extension to {\em Markov Modulated} input flows as well as Markovian server state modelling and prioritized service.

%Note that waiting times for heavy load seems to appear in reversed order, as compared to the lightly loaded case. This is an artifact stemming from the chosen statistic, where discarded packets are assigned zero waiting. In the lightly loaded case more packets are discarded, hence the result. A similar approach coudl yield expected waiting for only accepted packets. 

%&=&\frac{1}{\tau} \int_0^{\tau} E[I_{x(\eta) \in A}] d\eta \nonumber \\ 
%&=&\frac{1}{\tau} \int_0^{\tau} P(x(\eta) \in A) d\eta \nonumber \\
%&=&\frac{1}{\tau} \int_0^{\tau} P(x(\eta) \in A) d\eta \nonumber \\ 
%&=&\sum_{l=0}^M \frac{1}{\tau} \int_0^{\tau} P(x(\eta) \in A \ \wedge \ T(t)=l \ \wedge \ {\bf z} \neq \ \varepsilon) + P(x(\eta) \in A \ \wedge \ T(t)=l \ \wedge \ {\bf z} = \varepsilon) d\eta \label{time_cont1}
%\end{eqnarray} 

%and rely on the implementations in MatLab symbolic toolbox. To provide the reader with some sense of the practical feasibility we give computation times for a variety of problem scales.

%\frac{d}{dt} P^{T'}_t = P^{T'}_t {\bf \Gamma}^T 

\section{Appendix A}

Theorem: $Q(0)T(0)=0$ implies $Q(t)T(t)=0$ identically for all $t>0$.
\subsection{Proof}
First assume in \eqref{token} that $Q^-_n=0$, then $Q^+_n=0$. On the other hand, assume $Q^-_n>0$ and $T^-_n=0$. Then by the second equation of \eqref{token} $T^+_n=0$.\\
Next, assume in \eqref{arrival} that $Q(t^-)=0$. Then either $Q(t^+)=0$ or $T(t^-)=0$ by the two former equations of \eqref{token}. If $T(t^-)=0$, then $T(t^+)=0$ by the third equation of \eqref{arrival}. On the other hand assume $T(t^-)=0$ then directly $T(t^+)=0$ by the third equation of \eqref{arrival}.

\section{Appendix B}
For $K^-_n > 0$, yielding $Q^-_n > 0$ and $T^-_n=0$
\begin{equation}
K^+_n = K^-_n-1 \nonumber 
\end{equation}
and for $K^-_n = 0$, yielding $Q^-_n = T^-_n = 0$
\begin{equation}
K^+_n = \max\{-M,-1\}=-1 \nonumber
\end{equation}
and finally for $K^-_n < 0$, yielding $Q^-_n = 0$ and $T^-_n > 0$
\begin{equation}
K^+_n = \max\{-M,K^-_n-1\} \nonumber
\end{equation}
which aggregates to
\begin{equation}
K^+_n = \max\{-M,K^-_n-1\} \label{Ktoken_app}
\end{equation}

\section{Appendix C}
For $K(t^-) > 0$, yielding $Q(t^-) > 0$ and $T(t^-)=0$
\begin{equation}
K(t^+) = \min\{L,K(t^-)+1\} \nonumber 
\end{equation}
and for $K(t^-) = 0$, yielding $Q(t^-) = 0$ and $T(t^-)=0$
\begin{equation}
K(t^+) = \min\{L,K(t^-)+1\} \nonumber 
\end{equation}
and finally for $K(t^-) < 0$, yielding $Q(t^-) = 0$ and $T(t^-)>0$
\begin{equation}
K(t^+) = K(t^-)+1 \nonumber
\end{equation}
which aggregates to 
\begin{equation}
K(t^+) = \min\{L,K(t^-)+1\} \nonumber 
\end{equation}
yielding for the entire period $[(n-1)\tau, n\tau)$ 
\begin{equation}
K^-_{n} = \min\{L,K^+_{n-1}+a_n\} \label{Kentire_app} 
\end{equation}
Combining \eqref{Ktoken_app} and \eqref{Kentire_app} gives
\begin{equation}
K^+_n = \max\{-M,\min\{L,K^+_{n-1}+a_n\}-1\} \label{Kcomb:app}
\end{equation}

\section{Appendix D}
We define the function $B:Z \times {\mathbf N} \rightarrow {\mathbf N}$ by
\begin{equation}
B(l,L)=\# \{{\bf z} \in Z^* \ | \ |l {\bf z}| \leq L\} \nonumber
\end{equation} or less formally the number of buffer states with a buffer head $l$. 
We readily identify
\begin{equation}
R_L=\sum_{l \in Z} B(l,L)  \label{R_app} \nonumber
\end{equation}
and the recursion
\begin{equation}
B(l,L)=\sum_{v \in Z} B(v,L-l)  \label{rekurs}
\end{equation} 
as well as initial conditions
\begin{eqnarray}
B(l,L)&=&1 \ \ \mbox{for} \ \ l \leq L\leq l + \min(Z)  \nonumber \\
&=&0 \ \ \mbox{for} \ \ L < l  \nonumber
\end{eqnarray} 
We define the upper bound $\bar{B}(L)$ by
\begin{equation}
\bar{B}(L) = \max_{l \in Z} B(l,L) \label{uppbound}
\end{equation}
and immediately conclude that $\bar{B}(1) \leq 1$. Combining \eqref{rekurs} with \eqref{uppbound} yields
\begin{equation}
\bar{B}(L) = \max_{l \in Z} B(l,L)=\max_{l \in Z} \{\sum_{v \in Z} B(v,L-l)\} \label{uppbound2_app}
\end{equation} 
Next we identify $\alpha>1$ so that $\bar{B}(L) \leq \alpha^L$ and insert in \eqref{uppbound2}, i.e.
\begin{equation}
\bar{B}(L) \leq \max_{l \in Z} \{\sum_{v \in Z} \alpha^{L-l}\}=\#Z \max_{l \in Z} \alpha^{L-l}=\#Z \alpha^{L-\min\{ Z\}} \label{uppbound2}
\end{equation}
We proceed by finding $\beta$ such that 
\begin{equation}
\beta^L = \#Z \beta^{L-\min\{ Z\}} \label{beta}
\end{equation}
or
\begin{equation}
\beta^{\min\{ Z\}} = \#Z \Rightarrow \beta=\#Z^{\frac{1}{\min\{ Z\}}} \label{beta1}
\end{equation}
Finally it is readily proven recursively, that if $\bar{B}(1) \leq 1$
\begin{equation}
\bar{B}(L) \leq \beta^L = (\#Z^{\frac{1}{\min\{ Z\}}})^L \label{Bbar_app}
\end{equation}

%\begin{eqnarray}
%Q^+_n &=& \max\{0,min\{L,Q^+_{n-1}+a_n\}\} \nonumber \\
%T^+_n &=& T^+_n  \ \ \mbox{for} \ \ min\{L,Q^+_{n-1}+a_n\}>0 \nonumber \\
%T^+_n &=& \min\{M,T^+_n+1\}  \ \ \mbox{for} \ \ min\{L,Q^+_{n-1}+a_n\}=0 \nonumber 
%\end{eqnarray}

%Turner, J., New directions in communications (or which way to the information age?). Communications Magazine, IEEE 24 (10): 8–15. ISSN 0163-6804, 1986.
  %
%Andrew S. Tanenbaum, Computer Networks, Fourth Edition, ISBN 0-13-166836-6, Prentice Hall PTR, 2003., page 401.
%
%ATM Forum, The User Network Interface (UNI), v. 3.1, ISBN 0-13-393828-X, Prentice Hall PTR, 1995.
%
%ITU-T, Traffic control and congestion control in B ISDN, Recommendation I.371, International Telecommunication Union, 2004, Annex A, page 87.
%
%"Linux HTB Home Page". Retrieved 2013-11-30.

%Boudec

%R.J. McEliece, J. Murphy, M. Jennings and Z. Yu, "On Simplified Modelling of the Leaky Bucket", Proc. Of the IEE 13th UK IEE Teletraffic Symposium, Strathclyde, UK, 18-20 March 1996.
\bibliographystyle{IEEEtran}
\bibliography{literature}
%\begin{IEEEbiography}[{\includegraphics[width=1in,height=1.25in,clip,keepaspectratio]{103206.jpg}}]%
%{Henrik Schioler}
%serves as Associate Professor at the
%Department of Electronic Systems/Automation and
%Control at the Aalborg University. Research interest
%include performance evaluation of networked systems, queueing systems, network calculus and stochastic hybrid/switched dynamical systems.
%\end{IEEEbiography}
%\mbox{}\\
%\begin{IEEEbiography}[{\includegraphics[width=1in,height=1.25in,clip,keepaspectratio]{102987.jpg}}]%
%{John Josef Leth}
%serves as Associate Professor at the
%Department of Electronic Systems/Automation and
%Control at the Aalborg University. Research interest
%include (stochastic) hybrid/switched dynamical systems,
%as well as mathematical control theory, such as
%optimal control theory for nonlinear systems, realization theory for nonlinear systems, and infinite
%dimensional system theory.
%\end{IEEEbiography}
%\mbox{}\\
%\begin{IEEEbiography}[{\includegraphics[width=1in,height=1.25in,clip,keepaspectratio]{141143.png}}]%
%{Shibarchi Majumder}
%received B. Tech. in Aerospace Engineering and M. Tech. in Avionics Engineering in 2014. At present, he is working at the Department of Electronics Systems , Aalborg University, Denmark as a PhD researcher. His research interests include real-time embedded computing,  mixed-criticality partitioned systems and time-predictable on-chip networks. 
%
%\end{IEEEbiography}

\end{document}